\documentclass[12pt]{elsarticle}
\RequirePackage[left=0.8in,right=0.8in,top=1in,bottom=1in]{geometry}
\usepackage[utf8x]{inputenc}
\usepackage{graphicx}
\usepackage{epstopdf}
\usepackage{amsmath}
\usepackage{multirow}
\usepackage{subfigure}
\usepackage{amssymb}
\usepackage{booktabs}
\usepackage{setspace}
\usepackage{xcolor,colortbl}
\usepackage{nomencl} 
\makenomenclature
\graphicspath{ {fig/} }
\usepackage[compact]{titlesec}

\journal{Journal}
\begin{document} 
\begin{frontmatter}
\title{Assessing the practicability of the condition used for dynamic equilibrium in Pasinetti's theory of distribution.}
\author[rvt]{A. Jayakrishnan}
\author[els]{Anil Lal S\corref{cor1}}

\cortext[cor1]{Corresponding author, anillal@cet.ac.in}
\address[rvt]{Ielektron Technologies, IIT Madras Research Park, Taramani, Chennai, India - 600 113 \\ Email: a.jayakrishnan.tfmab@gmail.com }
\address[els]{College of Engineering,Trivandrum, Kerala, India - 695016. \\ Email: anillal@cet.ac.in}

%
%
\begin{abstract}
{\it In this note an assessment of the condition \(K_w/K=S_w/S\)  is made to interpret its  meaning to the  Passineti's theory of distribution\cite{pasinetti1962rate}. This condition leads the theory to enforce the result \(s_w\rightarrow0\) as \(P_w\rightarrow 0\),  which is the Pasinetti's description about behavior of the workers. We find that the Pasinetti's claim, of long run worker's propensity to save as not influencing the distribution of income between profits and the wage can not be generalized. This claim is found to be valid only when \(W>>P_w\) or \(P_w=0\) with \(W\ne0\).  In practice, the Pasinetti's  condition imposes a restriction on the actual savings by one of the agents to a lower level compared to its full saving capacity.  An implied relationship  between the propensities to save by workers and capitalists shows that the Passineti's condition can be practiced only through a contract for a constant value of \(R=s_w/s_c\), to be agreed upon  between the  workers and the capitalists. It is showed that the Passineti's condition can not be described as a dynamic equilibrium of economic growth. Implementation of this condition (a) may lead to accumulation of unsaved income, (b) reduces  growth of capital, (c)is  not practicable and (d) is not warranted.  We have also presented simple mathematical steps for the derivation of the Pasinetti's final equation compared to those presented in \cite{pasinetti1962rate}}.
\end{abstract}


\end{frontmatter}

\doublespacing
\section{Introduction}
Pasinetti's theory of distribution\cite{pasinetti1962rate} has introduced a correction to the logical slip in Kaldor's theory\cite{kaldor1955alternative} of not allocating profit on worker's saving.   Passineti argued that whenever a worker saves a part of his income, he must be allowed to own a portion of the total capital. Otherwise the workers do not save at all. The  ownership of assets generates a profit share to the workers in the same manner as that of capitalists. The distribution is expressed as
\begin{equation}
 P=P_c+P_w \label{eqn1}
\end{equation}
Where \(P_c\) and \(P_w\) are profit shares to capitalists and workers respectively. This correction leads to workers getting a feel of ownership along with a consequent  increase in their total income. Now, the total income of workers becomes  \(P_w+W\), where \(W\) is  worker's wage. The propensities to save by the workers (\(s_w\)) and the capitalists (\(s_c\)) result in the partial savings of capitalists \(S_c=s_cP_c\) and workers \(S_w=s_w(P_w+W)\) and the total saving \(S\) as
\begin{equation}
 S=S_c+S_w=s_cP_c+s_w(P_w+W) \label{eqn2}
\end{equation}
The purpose of this note is to interpret the meaning  of the following  condition Pasinetti has used for the dynamic equilibrium of economic growth.
\begin{equation}
 \frac{S_w}{S}=\frac{K_w}{K} \label{asump1}
\end{equation}
where \(K_w\), \(K_c\)  and \(K=K_w+K_c\) are the quantity of capital owned by workers, capital owned by the capitalists and total capital respectively. The condition in equation(\ref{asump1}) is referred to in this note as  Pasinetti's condition for the dynamic equilibrium. 
Another assumption used in the derivation of the  Pasinetti's final equation for the theory of distribution is a constant profit ratio given by
\begin{equation}
 \frac{P_w}{K_w}=\frac{P_c}{K_c}=\frac{P}{K} \label{asump2}
\end{equation}
Dividing equation(\ref{eqn1}) through out by \(K\) we can write
\begin{eqnarray}
 &&\frac{P}{K}-\frac{P_w}{K}=\frac{P_c}{K} \nonumber \\
 &&\frac{P}{K}-\frac{P_w}{K_w}\frac{K_w}{K}=\frac{P_c}{K} \nonumber \\
 &&\frac{P}{K}-\frac{P}{K}\frac{K_w}{K}=\frac{P_c}{K} ~~~(\because \mbox{constant profit ratio in equation-\ref{asump2}}) \nonumber\\
 &&\frac{P}{K}\left[1-\frac{K_w}{K}\right]=\frac{P_c}{K} 
 \end{eqnarray}
Now, the  Pasinetti's condition for dynamic equilibrium in equation(\ref{asump1}) is applied to simplify the above expression to obtain the following equation.
\begin{eqnarray}
 &&\frac{P}{K}\left[1-\frac{S_w}{S}\right]=\frac{P_c}{K} ~~~~~\mbox{or}~~~~~~ \frac{P}{K}\left[\frac{S_c}{S}\right]=\frac{P_c}{K} \label{step1}
 \end{eqnarray}
 By taking  \(S=I\) and \(S_c=s_cP_c\), we get
\begin{eqnarray}
 &&\frac{P}{K}\left[\frac{s_cP_c}{I}\right]=\frac{P_c}{K} \implies P_c\left[\frac{P}{K}-\frac{1}{s_c}\frac{I}{K}\right]=0 \nonumber
 \end{eqnarray}
In a capitalist economy \(P_c\ne0\) and hence we write  the  Pasinetti's final equation for distribution as 
\begin{equation}
 \frac{P}{K}=\frac{1}{s_c}\frac{I}{K} \label{pf1}
\end{equation}
It may be noted the mathematical steps used for obtaining the above final equation is simple which students of economics can more easily understand and comprehend in comparison to the one given in \cite{pasinetti1962rate}. Now we will look at the  difference between Pasinetti's model and Kaldor's model.  In Kaldor's theory \(P_w=0\), hence by equations(\ref{eqn1})\&(\ref{eqn2}), we have
\begin{equation}
 \frac{P}{K}=\frac{P_c}{K}=\frac{S-s_wW}{Ks_c}=\frac{1}{s_c}\frac{\left(I-s_wW\right)}{K} \label{kal1}
\end{equation}
The Kaldor's theory of distribution with \(s_w\) taken as equal to zero is
\begin{equation}
 \frac{P}{K}=\frac{1}{s_c}\frac{I}{K} \label{kal2}
\end{equation}
Pasinetti's theory as shown in equation(\ref{pf1}), obtained without making any assumption on the propensities to save by the workers and Kaldor's theory obtained with \(s_w=0\) as shown in equation(\ref{kal2}),  both seem to be exactly the same. Pasinetti has given an interpretation to  this match  between the equations as a long run insignificance of worker's propensities to save(\(s_w\rightarrow 0\)).  In this manner he claims that, in the long run worker's propensity to save does not influence the distribution of income between profits and wage. We argue that the comparison with the assumption \(s_w=0\) of Kaldor's model by  Pasinetti to arrive at a conclusion about the long run worker's propensities to save is not correct.  Truly,  Pasinetti's model  starts with a single difference from  Kaldor's model, which is in the share of profit to workers(\(P_w\)). In Kaldor's model \(P_w=0\), whereas it is greater than zero in Pasinetti's model. However,  difference between the two models widens by a larger extent, due to  inclusion of the Pasinetti's condition for dynamic equilibrium (equation-\ref{asump1}). We now demonstrate the differences between  Pasinetti's and Kaldor's models. For this, an alternate form of the Pasinetti's final equation is first derived by subtracting the LHS and RHS of equation(\ref{step1}) from \(P/K\) as in the following.
\begin{eqnarray}
 \frac{P}{K}-\frac{P}{K}\left[\frac{S_c}{S}\right]=\frac{P}{K}-\frac{P_c}{K} \implies \frac{P}{K}\frac{S_w}{S}=\frac{P_w}{K} \nonumber
\end{eqnarray}
Putting \(S_w=s_w(P_w+W)\) and taking \(S=I\), we get for \(P_w\ne0\) 
\begin{eqnarray}
 \frac{P}{K}=\frac{1}{s_w}\left(\frac{P_w}{W+P_w}\right)\frac{I}{K} \label{pf2}
\end{eqnarray}
By equating the right hand sides of equations(\ref{pf1}) and (\ref{pf2}), we get
\begin{eqnarray}
 s_c=s_w\left(1+\frac{W}{P_w}\right) \label{const1}
\end{eqnarray}
It shows that the application of the Pasinetti's condition results in the propensities getting constrained to satisfy  equation(\ref{const1}) and therefore the claim on insignificance of \(s_w\) does not remain generalized. The above equation can not be achieved  naturally, as the propensities  depend on the behavior the capitalists and the workers. Practically, the satisfaction of the condition  in equation(\ref{const1}) can be realized through a force of contract or agreement  between workers and capitalists. 
\section{Observations on the relationship between propensities to save}
Some observations and the meanings of the equation(\ref{const1}) between the propensities to save by workers and capitalists are enumerated in the following. 
\begin{enumerate}
\item In the limit as \(P_w\rightarrow 0\), the value of \(s_w\) should tend to zero,  so as to get a finite value for \(s_c\) as per equation(\ref{const1}). This enforces with confirmity, the idea that workers do not save at all if no portion of the profit is shared to them. This idea is incorporated in to the Pasinetti model by merely including the term \(P_w\) in equation(\ref{eqn2}), which does not enforce \(s_w\rightarrow 0\) as \(P_w\rightarrow 0\). Here  the logical correction by enforcing \(s_w\rightarrow 0\) as \(P_w\rightarrow 0\)
is resulted from the uses of the Passineti's condition and the constant profit ratio. 
\item Pasinetti's model is  silent  about the spending of \(W-C_w\) when \(P_w=0\), where \(C_w\) is the  worker's aggregate consumption. 
\item For fixed values of \(s_c\) and \(P_w\), an increase of worker's wage  leads to a reduction of \(s_w\). This is an inconsistency. Here, an increase of worker's income due to rise of worker's wages results in a lesser savings.
\item The  equations(\ref{pf1}) and (\ref{kal2}) are same only in terms the variables \(s_c\) and \(I\). But the other external variables involved in the two models occur with different ranges/proportions. Hence the interpretation made by Pasinetti on the external variable \(s_w\) as a consequence of the same form in equations(\ref{pf1}) and (\ref{kal2})  can not be generalized. 
\item The value of \(s_w\) as not influencing the distribution of income between profits and the wage is operational only in conditions such as \(P_w=0\) when \(W\ne0\) or \(W>>P_w\).
\end{enumerate}
The above observations related to interpreting the Pasinetti's final equation and relationship between propensities motivated to look into the insight of this condition for the dynamic equilibrium used by Pasinetti. 
\section{An insight into Pasinetti's condition}
 In the section we obtain a proof to show that the Pasinetti's condition can not represent a dynamic equilibrium. In the dynamic equilibrium of growth, savings translated into investments in each period is equal to the rate of capital growth in that period. Mathematically,
 \begin{eqnarray}
  S_w=\frac{dK_w}{dt};~~~S_c=\frac{dK_c}{dt}~~~\mbox{and}~~~S=\frac{dK}{dt}
 \end{eqnarray}
 Therefore
 \begin{eqnarray}
  K_w(t)=\int_0^tS_w dt;~~~~~K_c(t)=\int_0^tS_c dt~~~\mbox{and}~~~K(t)=\int_0^tS dt
 \end{eqnarray}
 And
 \begin{eqnarray}
  \frac{K_w}{K}=\frac{\int_0^tS_w dt}{\int_0^tS dt}
 \end{eqnarray}
As per the Pasinetti condition in equation(\ref{asump1}), this  ratio is equal to \(S_w/S\). This is possible  only if the instantaneous savings \(S_w(t)\) and \(S(t)\) satisfy the conditions
\begin{eqnarray}
 S_w(t)&=&Cf(t)  \\
 S_c(t)&=&Df(t)  \\
 S(t)&=&S_w+S_c=Ef(t)\nonumber
\end{eqnarray}
where \(E=C+D\). Clearly, at any point in time  the ratio of propensities is \(S_w/S_c=C/D=R\). This ratio rule has to be strictly followed at all the time periods without deviating. For keeping a constant value for \(R\), the actual amount saved by an agent(worker or capitalist) should be different from the amounts that an agent is willing to save based on his Full Capacity to Save(FCS). Let \(FCS_w(t)\) and   \(FCS_c(t)\) be the Full Capacity to Save by the workers and capitalists respectively.  In order to write a mathematical form for the actual savings by workers and capitalists for obtaining a constant contract ratio \(R=S_w/S_c\) corresponding to a certain capacity ratio \(R_1=FCS_w(t)/FCS_c(t)\), we define two conditional functions
\begin{eqnarray*}
 M_1=\max(R_1-R,0) \\
 M_2=\max(R-R_1,0)
\end{eqnarray*}
Now the actual  savings  by workers and capitalists restricted using \(R\)  can be written as
\begin{eqnarray}
 S_w(t)&=&S_w(FCS_w(t),FCS_c(t),R)=FCS_w(t)-FCS_c(t) M_1  \label{will1}\\
 S_c(t)&=&S_c(FCS_w(t),FCS_c(t),R)=\frac{FCS_w(t)}{M_2+R_1} \label{will2}
\end{eqnarray}
The selection of savings in the above manner leads to lesser saving by one of the agents, who has a higher capacity/willingness to save. The agent who has a higher capacity to save is forced to save less during the process of negotiating a constant  value for \(R\).  The unsaved capacity to save by workers and capitalists, denoted as \(US_w(t)\) and \(US_c(t)\) are given by
\begin{eqnarray}
US_w(t)&=& FCS_c(t)M_1 \\
US_c(t)&=&\frac{FCS_c(t)M_2}{M_2+R_1}
\end{eqnarray}
The value of \(US(t)=US_w(t)+US_c(t)\) is either greater than or equal to zero and it becomes equal to zero only when both \(M_1\) and \(M_2\) are equal to zero. Thus,  \(US(t)>0\) is  common, \(US(t)=0\) is only rarely possible and \(US(t)=0~\forall t\) is almost impossible.  The consequences of  \(US(t)>0\) are (a) reduction in the growth of capital due to lesser total investments in all the time periods, (b) reduction in total profit \(P\) and (c) reduction in profit share for one of the agents. Another important drawback arises is from the fact that, the unsaved capacity of an agent in a period will become a component in the Full Capacity to Save of that agent in the next period.  This  higher capacity to save due to the accumulated  unsaved amounts from the previous periods may be further restricted from saving in the subsequent periods  as per equations(\ref{will1}) and (\ref{will2}). This may be a matter of a continuous divergence of interests of the agents, rather than attainment of a dynamic equilibrium. In this way the Pasinetti's condition for dynamic equilibrium as given by equation(\ref{asump1}) is not correct. We consider that the contract between the workers and capitalists to practically implement the Pasinetti's condition encompasses losses  from different factors.  Again, using the condition of contract, we can derive the following
\begin{eqnarray}
 &&\frac{S_w}{S_c}=\frac{C}{D}=\frac{P_w}{P_c} \nonumber \\
 &&\frac{s_w(W+P_w)}{s_cP_c}=\frac{P_w}{P_c} \nonumber
\end{eqnarray}
For \(P_c\ne0\), the above equation implies
\begin{eqnarray}
 s_c=s_w\left(\frac{W+P_w}{P_w}\right) \label{ver1}
\end{eqnarray}
From the equations(\ref{const1}) and (\ref{ver1}), it is clear that the relationship between the propensities is resulted from a contract for a constant value of \(R\) to be agreed upon between the workers and the capitalists. 
\section{Conclusions}
An assessment of the practicability of the Pasinetti's condition  \((K_w/K=S_w/S)\) used for the dynamic equilibrium of economic growth in his theory of distribution is carried out and the following conclusions have been made.
\begin{enumerate}
\item Pasinetti's condition leads to enforcement of the worker's behavior given by \(s_w\rightarrow0\) when \(P_w\rightarrow 0\).
\item The Pasinetti model deviates to a larger extent from Kaldor's model due to the inclusion of the Pasinetti's condition. Hence the claim by Pasinetti on the long run worker's propensity to save as not influencing the distribution of income between profits can not be generalized.
His claim is found to be valid only when \(W>>P_w\) or \(P_w=0\) with \(W\ne0\).
\item Pasinetti's condition implies  a relationship between the propensities to save by workers and the capitalists. The relationship between the propensities can not be achieved naturally and requires the force of a contract for a constant value of  \(R\) to be agreed upon between the workers and the capitalists. 
 \item The implementation of Pasinetti's condition is found to restrict the actual savings by one of the agents to lower levels compared to its full capacity for saving.
\item This condition leads to a reduction of total savings and hence reduces the growth of capital. The amount of unsaved income tend to increase with time. So the Pasinetti's condition can not be described as a dynamic equilibrium of economic growth.
\item It is concluded that the Pasinetti's condition is not practicable and hence not warranted for describing the dynamics of growth of economies.
\end{enumerate}

\bibliography{ref.bib}

\begin{thebibliography}{1}
\expandafter\ifx\csname url\endcsname\relax
  \def\url#1{\texttt{#1}}\fi
\expandafter\ifx\csname urlprefix\endcsname\relax\def\urlprefix{URL }\fi
\expandafter\ifx\csname href\endcsname\relax
  \def\href#1#2{#2} \def\path#1{#1}\fi

\bibitem{pasinetti1962rate}
L.~L. Pasinetti, Rate of profit and income distribution in relation to the rate
  of economic growth, The Review of Economic Studies 29~(4) (1962) 267--279.

\bibitem{kaldor1955alternative}
N.~Kaldor, Alternative theories of distribution, The review of economic studies
  23~(2) (1955) 83--100.

\end{thebibliography}

~~\\
\appendix       
\section*{Appendix A: Illustrative examples  on the calculation of actual savings using equations(\ref{will1}) and (\ref{will2})}
~~\\
Example-1: Given   \(R=\frac{1}{5}\), \(FCS_w=5\) and \(FCS_c=20\). \\
Now, \(R_1=\frac{5}{20}=\frac{1}{4}\), \(M_1=\max(R_1-R,0)=\max\left(\frac{1}{20},0\right)=\frac{1}{20}\)\\
and \(M_2=\max(R-R_1,0)=\max\left(-\frac{1}{20},0\right)=0\) \\
and we have by equations(\ref{will1}) and (\ref{will2})
\begin{eqnarray*}
 S_w(t)&=&FCS_w(t)-FCS_c(t) M_1=5-20\times \frac{1}{20}=4  \\
 S_c(t)&=&\frac{FCS_w(t)}{M_2+R_1}= \frac{5}{0+\frac{1}{4}}=20
\end{eqnarray*}
In this case the workers will save only \(4\) units which is less than their full capacity by keeping \(R=\frac{4}{20}=\frac{1}{5}\) as the required value. \\
~~~\\
Example-2: Given   \(R=\frac{1}{4}\), \(FCS_w=4\) and \(FCS_c=20\). \\
Now, \(R_1=\frac{4}{20}=\frac{1}{5}\), \(M_1=\max(R_1-R,0)=\max\left(-\frac{1}{20},0\right)=0\)\\
and \(M_2=\max(R-R_1,0)=\max\left(\frac{1}{20},0\right)=\frac{1}{20}\)\\
and we have by equations(\ref{will1}) and (\ref{will2})
\begin{eqnarray*}
 S_w(t)&=&FCS_w(t)-FCS_c(t) M_1=4-20\times 0=4  \\
 S_c(t)&=&\frac{FCS_w(t)}{M_2+R_1}= \frac{4}{\frac{1}{20}+\frac{1}{5}}=16
\end{eqnarray*}
In this case the capitalists will save only \(16\) units which is less than their full capacity by keeping \(R=\frac{4}{16}=\frac{1}{4}\) as the required value.

\end{document}